\documentclass[]{jpsj3}
\usepackage{txfonts}
\usepackage{comment}
\usepackage{txfonts}
\usepackage{braket,ulem,bm}
\usepackage[dvipdfmx]{color}
\usepackage{booktabs}

\title{Anomalous Quantum Transport Properties in Semimetallic Black Phosphorus}

\author{Kazuto~Akiba$^1$\thanks{E-mail: k{\_}akiba@issp.u-tokyo.ac.jp}, Astushi~Miyake$^1$, Yuichi~Akahama$^2$, Kazuyuki~Matsubayashi$^1$, Yoshiya~Uwatoko$^1$,
Hayato~Arai$^3$, Yuki Fuseya$^3$, and Masashi~Tokunaga$^1$}
\inst{$^1$The Institute for Solid State Physics, University of Tokyo, Kashiwa, Chiba 277-8581, Japan \\
$^2$Graduate School of Material Science, University of Hyogo, Kamigori, Hyogo 678-1297, Japan\\
$^3$Department of Engineering Science, University of Electro-Communications, Chofu, Tokyo 182-8585, 
Japan
}

\abst{
Magnetoresistance in single crystals of black phosphorus is studied at ambient and hydrostatic pressures.
In the semiconducting states at pressures below 0.71 GPa, the magnetoresistance shows periodic oscillations, 
which can be ascribed to the magneto-phonon resonance that is characteristic of high mobility semiconductors. 
In the metallic state above 1.64 GPa, the both transverse and longitudinal magnetoresistance show titanic increase
with exhibiting superposed Shubnikov-de Haas oscillations. 
The observed small Fermi surfaces, high mobilities and light effective masses of carriers in semimetallic black phosphorus are comparable 
to those in the representative elemental semimetals of bismuth and graphite.
}

\begin{document}
\maketitle

\begin{figure}
\begin{center}
\includegraphics[width=8.5cm]{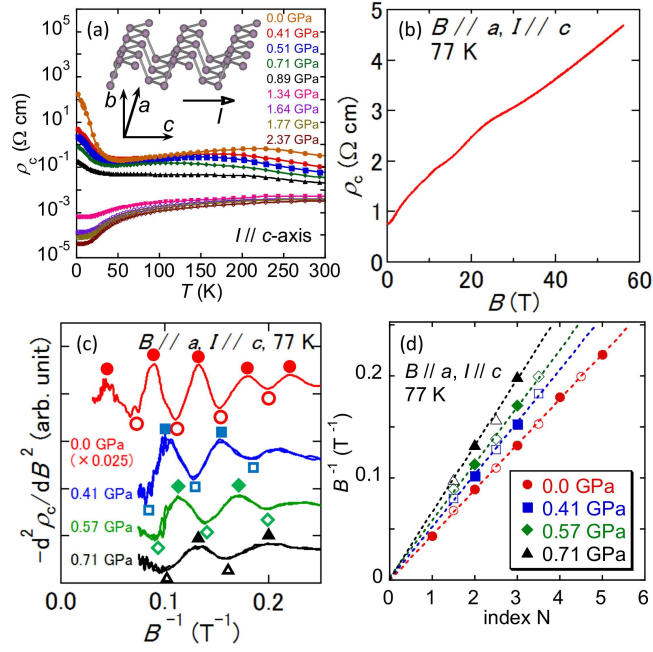}
\caption{(Color online)
(a) Temperature dependence of $\rho_\mathrm{c}$ at various pressures from 0.0 to 2.37 GPa. 
The inset illustrates the relation between the crystal structure and applied currents. 
(b) Magnetic field dependence of $\rho_\mathrm{c}$ at ambient pressure and temperature of 77 K. 
Magnetic field up to 56 T was applied along the $a$-axis.
(c) Second derivative of $\rho_\mathrm{c}$ as a function of $B^{-1}$ at ambient and hydrostatic pressures.
Peak and dip structures at each pressure are denoted by solid and open symbols, respectively. 
(d) Relation between inversed peak/dip fields $B^{-1}$ and the
index N in Eq. (\ref{eq_res}) (see text). 
\label{fig_1}}
\end{center}
\end{figure}

Black phosphorus (BP) is one of the stable allotropes of phosphorus at room temperature 
and consists of puckered honeycomb layers of phosphorus atoms in the $ac$-plane of the orthorhombic 
crystal\cite{Hultgren}. 
As shown in the inset of Fig. \ref{fig_1}(a), the directions along the zigzag chain 
and its normal correspond to the $a$- and $c$-axis, respectively. 
The monolayer of BP (called phosphorene) is known as a high mobility semiconductor
having a direct gap of 2 eV at the $\Gamma$ point
and attracts recent attention as a
new candidate for electrical devices next to graphene\cite{Li}.
In bulk BP, the honeycomb planes stack along the $b$-axis and are weakly bonded with the van der Waals force
with alternatively shifting along the $c$-axis, i.e., forming ABAB$\cdots$ stacks.
The interlayer interaction involves band dispersion along the $\Gamma$-$Z$ line.
As a result, the bulk BP becomes a narrow gap semiconductor with the direct gap of 300 meV 
at the $Z$ point\cite{Takao, Asahina}.
Application of hydrostatic pressure to BP further increases the interlayer interaction 
and then closes the gap
at about 1.5 GPa. 
This pressure-induced semiconductor-semimetal (SC-SM) transition has 
been experimentally observed in the transport and the
infrared absorption measurements under pressure\cite{Okazima, Akahama, Akahama_1986}.
Although the semimetallic state in this high mobility material is expected to exhibit unusual 
quantum transport properties, the details have not been revealed.
This motivates us to carry out the magnetoresistance measurements in BP under pressure. 

Single crystals of BP were synthesized under high pressure\cite{Endo_sample}. 
Transverse and longitudinal magnetoresistance were measured by the four-probe method on samples with the typical dimensions of 1.5 $\times$ 1.0 $\times$ 0.1 mm$^3$. 
Gold wires of 30 $\mu$m-thick were attached to the cleaved surfaces of the samples with MRX-713J carbon epoxy.
High pressure environments up to 2.4 GPa were realized by a piston cylinder-type pressure cell. 
The pressures in the sample space were determined by monitoring the superconducting transition temperatures of Pb set together with the samples. Glycerin or Daphne 7373 were used as pressure media.
For magnetoresistance measurements at ambient pressure, we utilized non-destructive 56-T magnets at The Institute for Solid State 
Physics at The University of Tokyo with the duration of about 36 ms, whereas the measurements under pressure were carried out up to 14 T with using superconducting magnets of Physical Property Measurement System (Quantum Design, Inc.).
Magnetic fields are applied along the $a$-axis throughout the entire measurements.

Figure \ref{fig_1}(a) shows the temperature dependence of the resistivity along the $c$-axis
($\rho_\mathrm{c}$) at various pressures.
Above 1.34 GPa, $\rho_\mathrm{c}$ shows metallic behavior 
in the whole temperature range between 2 and 300 K.
First, we focus on the transport properties of the semiconducting BP 
under ambient and relatively low pressures. 
Figure \ref{fig_1}(b) shows the field dependence of $\rho_\mathrm{c}$ 
at 77 K with a magnetic field ($B$) up to 
56 T applied along the $a$-axis. 
The $\rho_\mathrm{c}$ shows positive magnetoresistance accompanied by 
some hump structures below 30 T.
These structures are periodic in terms of $B^{-1}$ 
as shown by the solid and open circles in Fig. \ref{fig_1}(c). 
This periodic structure has been ascribed to the magneto-phonon resonance (MPR)\cite{Strutz}. 
The MPR is the resonant scattering of carriers by optical phonons and observed
as enhancement of resistance when the following resonance condition is fulfilled:
\begin{equation}
\hbar \omega_0=\mathrm{N}\hbar \omega_\mathrm{c}\,\,\,\,\,(\mathrm{N}=1,\, 2,\, 3,\dots)
\label{eq_res}
\end{equation}
where $\hbar \omega_0$ is the energy of the optical phonon and 
$\omega_\mathrm{c}=eB/m^*$ is the cyclotron frequency\cite{Nicholas}.
The $e$ and $m^*$ represent 
elemental charge and effective mass of carriers, respectively. 
The MPR 
becomes hardly visible at low temperatures because the thermal excitation of the optical phonon is suppressed.
Equation (\ref{eq_res}) can be transformed to the relation between inversed field and the integer index as
\begin{equation}
\dfrac{1}{B}= \dfrac{e}{m^*\omega_0}\mathrm{N}.
\label{eq_MPR_linefit}
\end{equation}
With using $\omega_0=2.60 \times 10^{13}$ s$^{-1}$ (138 cm$^{-1}$), which is the energy of the 
longitudinal optical (LO) 
phonon determined by the Raman scattering\cite{Shunji}, 
$m^*$ estimated from Eq. (\ref{eq_MPR_linefit}) is $0.153$ $m_0$ at ambient pressure,
which is consistent with the previously reported values of 0.146 $m_0$ at 20 K\cite{Narita} and 0.166 $m_0$
at 30 K\cite{Takeyama} determined by experiments of the cyclotron resonance.
Observation of the MPR indicates the existence of well-defined Landau levels in BP, i.e., 
$\omega_\mathrm{c}\tau = \mu B >1$ above 4 T.
Here, $\tau$ and $\mu$ denote the relaxation time and mobility of carriers, respectively.
Therefore, $\mu > 0.25$ m$^2$ V$^{-1}$ s$^{-1}$ $=$ 2500 cm$^2$ V$^{-1}$ s$^{-1}$
at 77 K, which is consistent with the earlier report\cite{Keyes, Akahama_1983}. 
Figure \ref{fig_1}(c) shows the pressure 
dependence of the second derivative $-$d$^2\rho_\mathrm{c}/$d$B^2$ at 77 K. 
The period of oscillation becomes long with applying 
pressure, which is interpreted as the increase of the coefficient $e/(m^*\omega_0)$ in Eq. (\ref{eq_MPR_linefit}). 
Assuming the relevant $\omega_0$ is insensitive to the 
pressure as the other phonon modes are\cite{Vanderborgh}, we can evaluate the pressure dependence of $m^*$ from the slopes of Fig. \ref{fig_1}(d). 
In Fig. \ref{fig_1}(d), half-integer indices correspond to the local minima in Fig. \ref{fig_1}(c).
Figure \ref{fig_1}(d) clearly indicates the reduction of $m^*$ by applied pressure.  
According to the $\bm{k}\cdot \bm{p}$ theory, the energy gap is roughly proportional to $m^*$ 
in the case of direct band gap semiconductors.
Hence, this implies the suppression of the energy gap by pressure.

\begin{figure}
\begin{center}
\includegraphics[width=8.5cm]{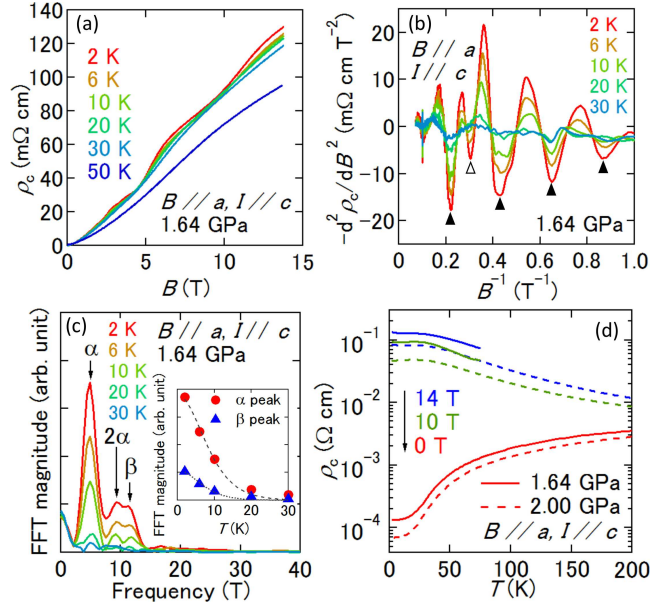}
\caption{(Color online)
(a) Transverse magnetoresistance at 1.64 GPa.
(b) Second derivative of $\rho_\mathrm{c}$ shown in (a). Solid and open arrowheads denote
the oscillatory structures originate from different cross-section of the Fermi surface.
(c) Fast Fourier transform (FFT) of SdH oscillation shown in (b). Detected frequency
components ($\alpha$, $\beta$, and $2\alpha$) are denoted by solid arrows. The inset shows the temperature
dependence of FFT magnitude at $\alpha$ and $\beta$. Dashed lines are fitted curves with using
Eq. (\ref{eq_LKamp}). 
(d) The temperature dependence of $\rho_\mathrm{c}$ in several magnetic fields at 1.64 (solid lines)
and 2.00 GPa (broken lines).
\label{fig_2}} 
\end{center}
\end{figure}

Now, we present the detailed transport properties of 
semimetallic BP.
Figure \ref{fig_2}(a) illustrates $\rho_\mathrm{c}$ at 1.64 GPa as a function of magnetic field
along the $a$-axis.
At 2 K, 
the magnetoresistance normalized by the value at zero-field reaches 
$\rho_\mathrm{c}(B)/\rho_\mathrm{c}(0)\sim 1000$ at 14 T.
Such a drastic change in magnetoresistance has been known in bismuth\cite{Kapitza}
and graphite\cite{McC},
typical elemental semimetals, and attracts
renewed interests in the other semimetals\cite{Ali}.

In addition to this approximately linear magnetoresistance, we can identify the superposed 
modulation.
The modulated components are clearly visible in their second derivatives shown in Fig. \ref{fig_2}(b).
Contrary to the MPR in the semiconducting state, the oscillating component periodic to $B^{-1}$ grows
up as the temperature decreases, and hence, can be ascribed to the Shubnikov-de Haas
(SdH) oscillations.
We can see a major frequency component (solid arrowheads) and additional small structures 
(an open arrowhead) 
in Fig. \ref{fig_2}(b), suggesting there are at least two frequency components. 
Frequency spectrum and its temperature dependence are shown in Fig. \ref{fig_2}(c).
We can see two distinct peaks, $\alpha$ (4.9 T) and $\beta$ (11 T), and 
the second harmonic $2\alpha$ (9.8 T) as marked by arrows.
Assuming the semimetallic state, there should be at least two Fermi pockets, 
i.e., each one electron- and hole-pocket. 
The cross-sectional areas of each Fermi surface are 0.47 $\times 10^{-3}$\AA$^{-2}$ and 1.0 $\times 10^{-3}$\AA$^{-2}$ for $\alpha$ and $\beta$, which is approximately
0.017 and 0.037 \% of that of the first Brillouin zone at the $k_z=0$ plane. 
For spherical Fermi surfaces, 
these small pockets roughly correspond to the carrier densities of 10$^{16}$ cm$^{-3}$.
Although the actual Fermi surface should be highly anisotropic\cite{Narita, Takeyama},
its carrier density can be comparable to those of bismuth and graphite.

In the semiclassical treatment, the SdH oscillations are represented by 
Lifshitz-Kosevich (LK) formula\cite{LK_form, LK_note}. 
In this formula, the thermal damping factor of amplitude $R_\mathrm{T}$ is given as follows:
\begin{equation}
R_\mathrm{T}= \dfrac{2\pi^2 k_\mathrm{B}T/\hbar \omega_\mathrm{c}}{\sinh(2\pi^2 k_\mathrm{B}T/\hbar \omega_\mathrm{c})},
\label{eq_LKamp}
\end{equation}
where $k_\mathrm{B}$ is the Boltzmann constant.
The observed temperature dependence of the oscillation amplitude 
can be reproduced by Eq. (\ref{eq_LKamp}) with $m^*$ $\sim$ 0.02 $m_0$
for both $\alpha$ and $\beta$ as shown in the inset of Fig. \ref{fig_2}(c).

\begin{figure}
\begin{center}
\includegraphics[width=8.5cm]{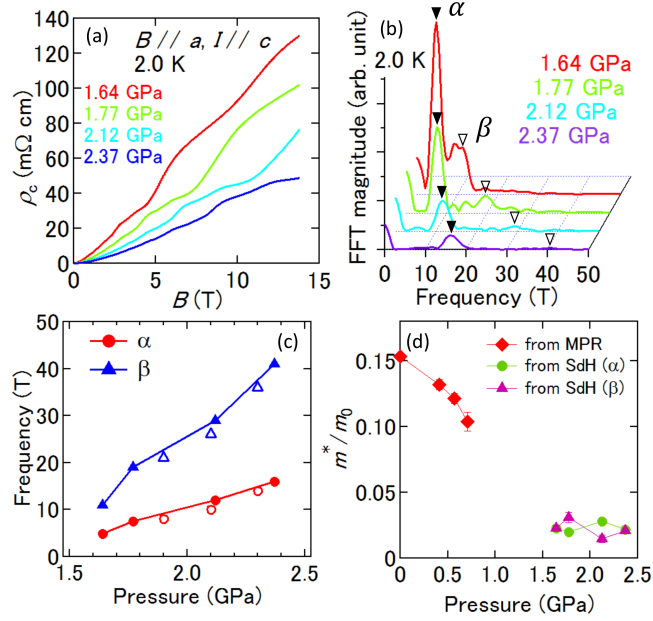}
\caption{(Color online)
(a) Magnetic field dependence of $\rho_\mathrm{c}$ at various pressures. 
(b) FFT spectrum of the SdH oscillations under several pressures. 
The $\alpha$ and $\beta$ peaks are traced by solid and open arrowheads, respectively. 
(c) Pressure dependence of the $\alpha$ and $\beta$ peaks.
Open symbols indicate the result on another sample piece.
(d) Pressure dependence of the effective mass estimated from MPR in semiconducting region
( $<$ 1.5 GPa) and from the temperature dependence of SdH amplitude in semimetallic region ( $>$ 1.5 GPa).
Vertical errorbars represent the statistical error in curve fitting. 
\label{fig_3}} 
\end{center}
\end{figure}
 
Next, we demonstrate the magnetoresistance at various pressures in the semimetallic state. 
Figure \ref{fig_3}(a) shows $\rho_\mathrm{c}$ at 2.0 K
as a function of applied field along the $a$-axis.
As illustrated in the FFT spectra [Fig. \ref{fig_3}(b)], the peaks $\alpha$ and $\beta$ 
detected at 1.64 GPa move toward 
higher frequency, indicating the enlargement of the Fermi surfaces by pressure. 
We investigated the pressure dependence of these peaks in another sample
as shown by the open symbols in Fig. \ref{fig_3}(c)
and obtained consistent results.
The pressure dependence of $m^*$ determined through analyses of the MPR and the SdH 
is shown in Fig. \ref{fig_3}(d).
$m^*$ of semimetallic region is significantly smaller than 
that of semiconducting region. 

At the vicinity of the SC-SM transition pressure, 
the quantum limit state in which all the carriers are accommodated in the lowest Landau subband
can be realized at around 10 T.
Recent experimental studies on graphite suggested the emergence of an excitonic phase
at the vicinity of the quantum limit state in high magnetic fields\cite{Fauque, Akiba}.
To clarify whether an exotic phase emerges in BP, we studied
the temperature dependence of the resistivity as shown in Fig. \ref{fig_2}(d).
The temperature dependence at zero-field is metallic as mentioned above, while in magnetic fields, 
they turn to the semiconducting ones in temperature up to 200 K.
Such apparent semiconducting-like behavior in semimetals can be explained 
by simple two carrier model\cite{Du}.
In the semiclassical approach, the transverse resistivity ($\rho_\mathrm{\perp}$), 
is expressed as\cite{Ziman}
\begin{equation}
\rho_\mathrm{\perp}=\dfrac{\rho_\mathrm{e}\rho_\mathrm{h}(\rho_\mathrm{e}+\rho_\mathrm{h})+(\rho_\mathrm{e}R_\mathrm{h}^2+\rho_\mathrm{h} R_\mathrm{e}^2) B^2}{(\rho_\mathrm{e}+\rho_\mathrm{h})^2+(R_\mathrm{e}+R_\mathrm{h})^2B^2},\label{eq_Ziman}
\end{equation} 
where $R_\mathrm{e,h}=1/(n_\mathrm{e,h} q_\mathrm{e,h})$ is the Hall coefficients 
of electron and hole carriers
($n_\mathrm{e,h}$ and $q_\mathrm{e,h}$ represent the density and the charge of 
electron and hole carriers, respectively)
and $\rho_\mathrm{e,h}$ is the resistivity of 
each carrier at zero-field.
In a compensated semimetal, we can set $-q_\mathrm{e}=q_\mathrm{h}$ 
and $n_\mathrm{e}=n_\mathrm{h}$, i.e., $-R_\mathrm{e}=R_\mathrm{h}=R$.
Assuming
$\rho_\mathrm{e}(T)=\rho_\mathrm{h}(T)=\rho_0(T)$ at zero-field, Eq. (\ref{eq_Ziman}) is reduced to 
\begin{equation}
\rho_\mathrm{\perp}=\dfrac{\rho_0(T)}{2}+\dfrac{R^2B^2}{2\rho_0(T)}.\label{eq_twocarrier}
\end{equation}
In low-carrier systems,
the second term in Eq. (\ref{eq_twocarrier}) becomes dominant, and hence $\rho_\mathrm{\perp}$ can show
the superficial insulating-like behavior even in a metallic sample. 

\begin{figure}
\begin{center}
\includegraphics[width=8.5cm]{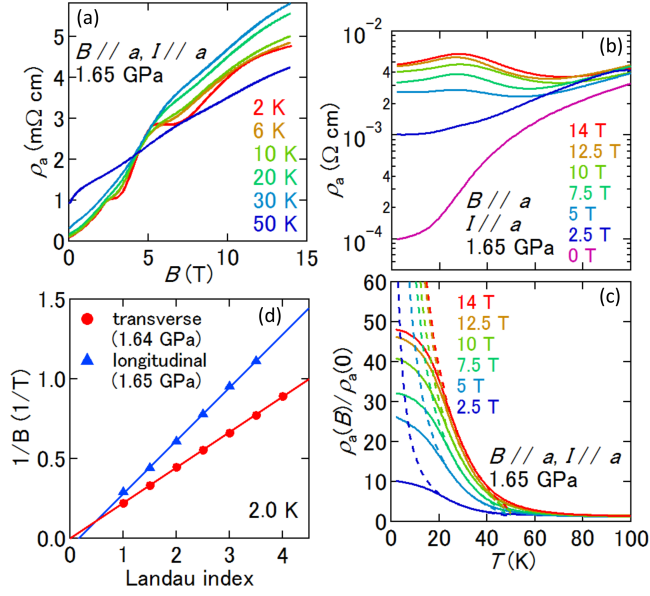}
\caption{(Color online)
(a) SdH oscillation in the longitudinal resistance $\rho_\mathrm{a}$ at 1.65 GPa.
(b) Temperature dependence of $\rho_\mathrm{a}$ at 1.65 GPa under several magnetic fields.
Horizontal axis is common to (c).
(c) Temperature dependence of $\rho_\mathrm{a}$ normalized by the resistivity at zero field $\rho_\mathrm{a}(0)$ at 1.65 GPa. Fitting curves based on Eq. (\ref{eq_Roth_fit}) in the text are shown by broken lines.
(d) Fan diagram at $\alpha$ peak for transverse ($B$ $\perp$ $I$) and longitudinal ($B$ $\parallel$ $I$) configurations.  
\label{fig_4}}
\end{center}
\end{figure}
To clarify the intrinsic features of $\rho_\mathrm{e}(T)$ or $\rho_\mathrm{h} (T)$, 
it is rather suitable to study longitudinal magnetoresistance.
Figure \ref{fig_4}(a) shows $a$-axis resistivity ($\rho_\mathrm{a}$) as a function of $B$
applied parallel to the $a$-axis at pressure of 1.65 GPa.
In this longitudinal geometry, we clearly observed SdH oscillations with the frequencies almost identical to those in the transverse one at 1.64 GPa. Thereby, quantum limit state is achieved at fields above 11 T. Even in this longitudinal geometry, we observed non-saturating significant positive magnetoresistance. At 2 K, the ratio $\rho_\mathrm{a}(B)/\rho_\mathrm{a}(0)$ reaches approximately 50 at 14 T.

As shown in Fig. \ref{fig_4}(b), 
this longitudinal magnetoresistance shows peak structure at around 30 K as a function of temperature. Such non-monotonic behavior, however, may not be ascribed to the emergence of a novel phase below this temperature. 
Figure \ref{fig_4}(c) shows the temperature dependence of the $\rho_\mathrm{a}(B)$
normalized by the value at zero field $\rho_\mathrm{a}(0)$.
The traces show monotonic behaviors at all the fields. 
According to a transport theory in the quantum limit\cite{RA},
The longitudinal resistivity 
$\rho_\mathrm{\parallel}(B,\,T)/\rho_\mathrm{\parallel}(0,\,T)$
for the classical statistics, in which the Fermi energy is smaller than $k_\mathrm{B}T$,
is represented as,
\begin{equation}
\dfrac{\rho_\mathrm{\parallel}(B,\,T)}{\rho_\mathrm{\parallel}(0,\,T)}
= \dfrac{1}{3}\left( \dfrac{\hbar \omega_\mathrm{c}}{k_\mathrm{B}T} \right)
= \dfrac{\hbar e}{3k_\mathrm{B}m^*}\dfrac{B}{T}
\label{eq_Roth_fit}
\end{equation} 
in case that the scattering mechanism is dominated by $\delta$-function impurity potential 
or acoustic phonon.
As shown with broken lines in Fig. \ref{fig_4}(c) , $\rho_\mathrm{a}(B,\,T)/\rho_\mathrm{a}(0,\,T)$ 
is reasonably reproduced by Eq. (\ref{eq_Roth_fit}) at all the field by adjusting single parameter 
$m^*=0.005$ $m_0$ above 20 K where the classical statistics likely being relevant.
Although the $m^*$ value used in this analysis is different from 
that estimated above, this discrepancy is not unusual 
with taking into account the crudeness of the model.
Therefore, unusual temperature dependence of the transverse and longitudinal magnetoresistance in semimetallic BP does not show the emergence of a gapped phase at high magnetic field, but can be simply understood as characteristic behavior in a high mobility and low carrier-density semimetal comparable to bismuth and graphite.

Here, let us make a comment on the phase of the SdH oscillations. 
In Fig. \ref{fig_4}(d), we plot the relation between inversed dip fields of $\rho_\mathrm{a}$ 
and $\rho_\mathrm{c}$ and the Landau index.
Similar analyses are frequently utilized in topological materials to evaluate the Berry phases
from the values of the horizontal intercept in this diagram.
The intercept for $\rho_\mathrm{c}$ is 0.00 $\pm$ 0.01 similar to two-dimensional Dirac system,
whereas that for $\rho_\mathrm{a}$ is 0.15 $\pm$ 0.02 even for the same field direction.
In addition, we note that such analyses on resistivity make sense only in the case that Hall conductivity 
is much larger than the diagonal one\cite{Ando}.
Therefore, we cannot make reliable argument from the analyses of this diagram.

\begin{figure}
\begin{center}
\includegraphics[width=8.5cm]{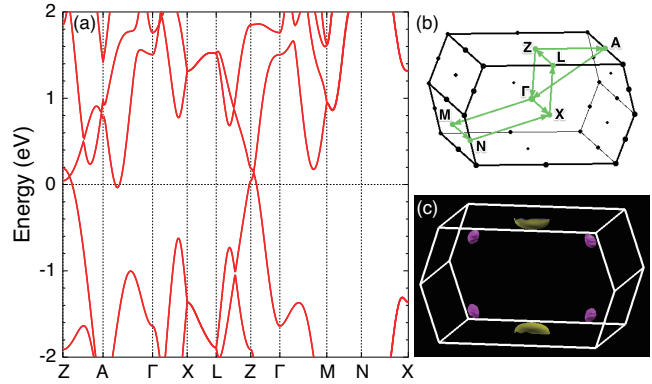}
\caption{(Color online) (a) Band structure at pressure of 2.0 GPa. (b) Brillouin zone and some of the symmetric points. (c) Fermi surfaces: one hole pocket at around the $Z$ point and four electron pockets between the $\Gamma$-$A$ line. The $b$-axis corresponds to the $\Gamma$-$Z$ direction and the $c$-axis does to $\Gamma$-$X$ direction.}
\label{fig_5}
\end{center}
\end{figure}

Finally, we show the results of band calculation under pressure and compare 
them with the experimental results.
Figure \ref{fig_5} (a) shows the band structure at pressure of 2.0 GPa by the first-principle calculations within the generalized gradient approximation using the OpenMX code\cite{OpenMX}. Two and one primitive obitals for $s$- and $p$-obritals are used for the pseudo-atomic orbitals basis functions, respectively. For the crystal structure parameters at 2.0 GPa, we linearly extrapolated the values at ambient pressure and 0.8 GPa of Ref. \citen{Asahina}. (We have checked that we can obtain almost the same results as Fig. \ref{fig_5} (a) even by using the crystal structure parameters given in Ref. \citen{Cartz}.)

At ambient pressure, the obtained band structure (not shown) is consistent with previous works\cite{Asahina,Qiao} and the direct band gap at the $Z$ point is 130 meV. At pressure of 2.0 GPa, the band overlap between the conduction and valance bands at the $Z$ point is 140 meV. There are one hole pocket at around the $Z$ point and four electron pockets locating between the $\Gamma$-$A$ line. (Their energy minimum are slightly off the $\Gamma$-$A$ line.) The hole pocket is highly anisotropic: the ratio of Fermi wavevectors is $k_{\rm F}^{Z-A}: k_{\rm F}^{Z-\Gamma}:k_{\rm F}^{Z-L}=21:13:3$, while the electron pockets are rather isotropic. Although the volume of the hole pocket should be four times larger than that of the electron pockets, the cross-sectional area of hole pocket is not necessarily so due to the highly anisotropic hole pocket. (The $a$-axis corresponds to the $Z$-$A$ direction.)

At the final stage in preparing this manuscript, we became aware of the paper
has been submitted to arXiv that reports similar quantum transport in semimetallic BP\cite{Xiang}.

In conclusion, we investigated the magnetoresistance in semiconducting and semimetallic black phosphorus under 
hydrostatic pressures. 
In the semiconducting state, we observed the magneto-phonon resonance
that is characteristic of high mobility materials.
In the semimetallc state above 1.6 GPa, we observed clear Shubnikov-de Haas oscillations 
and positive giant magnetoresistance effects. 
The analysis of the oscillation revealed its extremely small carrier density 
and light effective mass, which suggests that
black phosphorus can be a novel material to investigate the exotic quantum physics 
expected in the quantum limit state.


\begin{thebibliography}{99}
\bibitem{Hultgren} R. Hultgren, N. S. Gingrich, and B. E. Warren, J. Chem. Phys. \textbf{3,} 351 (1935).
\bibitem{Li} L. Li, Y. Yu, G. J. Ye, Q. Ge, X. Ou, H. Wu, D. Feng, X. H. Chen, and Y. Zhang, 
Nat. Nanotech. \textbf{9,} 372 (2014).
\bibitem{Takao} Y. Takao, H. Asahina, and A. Morita, J. Phys. Soc. Jpn. \textbf{50,} 3362 (1981).
\bibitem{Asahina} H. Ashahina, K. Shindo, and A. Morita, J. Phys. Soc. Jpn. \textbf{51,} 1193 (1982).
\bibitem{Okazima} M. Okazima, S. Endo, Y. Akahama, and S. Narita, Jpn. J. Appl. Phys. \textbf{23,} 15 (1984).
\bibitem{Akahama} Y. Akahama and H. Kawamura, Phys. Status Solidi B \textbf{223,} 349 (2001).
\bibitem{Akahama_1986} Y. Akahama, S. Endo, and S. Narita, Physica B \textbf{139-140,} (1986).
\bibitem{Endo_sample} S. Endo, Y. Akahama, S. Terada, and S. Narita, J. Appl. Phys. \textbf{21,} L482 (1982).
\bibitem{Strutz} T. Strutz, N. Miura, and Y. Akahama, Physica B \textbf{201,} 387 (1994).
\bibitem{Nicholas} R. J. Nicholas, Prog. Quant. Electr. \textbf{10,} 1 (1985).
\bibitem{Shunji} S. Sugai and I. Shirotani, Solid State Commun. \textbf{53,} 753 (1985).
\bibitem{Narita} S. Narita, S. Terada, S. Mori, K. Muro, Y. Akahama, and S. Endo, J. Phys. Soc. Jpn. \textbf{52,} 
3544 (1983).
\bibitem{Takeyama} S. Takeyama, N. Miura, Y. Akahama, and S. Endo, J. Phys. Soc. Jpn. \textbf{59,} 2400 (1989).
\bibitem{Keyes} R. W. Keyes, Phys. Rev. \textbf{92,} 580 (1953).
\bibitem{Akahama_1983} Y. Akahama, S. Endo, and S. Narita, J. Phys. Soc. Jpn. \textbf{52,} 2148 (1983).
\bibitem{Vanderborgh} C. A. Vanderborgh and D. Schiferl, Phys. Rev. B \textbf{40,} 9595 (1989).
\bibitem{Kapitza} P. Kapitza, Proc. R. Soc. London, Ser. A, \textbf{119,} 358 (1928).
\bibitem{McC} J. W. McClure and W. J. Spry, Phys. Rev. {\bf 165,} 809 (1968).
\bibitem{Ali} M. N. Ali, J. Xiong, S. Flynn, J. Tao, Q. D. Gibson, L. M. Schoop, T. Liang, N. Haldolaarachchige, M. Hirschberger, N. P. Ong, and R. J. Cava, Nature \textbf{514,} 205 (2014).
\bibitem{LK_form} D. Shoenberg, \textit{Magnetic oscillations in metals} (Cambridge University Press, 
Cambridge, 1984).
\bibitem{LK_note} LK formula is delived from semiclassical approximation, and hence, 
might be not appropriate to use in the vicinity of the quantum limit state.
We, however, use this formula because the similar analysis works well 
in the SdH oscillation in graphite\cite{Soule}.
\bibitem{Soule} D. E. Soule, J. W. McClure, and L. B. Smith, Phys. Rev. \textbf{134,} 2A (1963).
\bibitem{Fauque} B. Fauqu\'e, D. LeBoeuf, B. Vignolle, M. Nardone, C. Proust, and K. Behnia, Phys. Rev. Lett. \textbf{110,} 266601 (2013).
\bibitem{Akiba} K. Akiba, A. Miyake, H. Yaguchi, A. Matsuo, K. Kindo, and M. Tokunaga, J. Phys. Soc. Jpn.
\textbf{84,} 054709 (2015).
\bibitem{Du} X. Du, S. Tsai, D. L. Maslov, and A. Hebard, Phys. Rev. Lett. \textbf{94,} 166601 (2005).
\bibitem{Ziman} J. M. Ziman, \textit{Principles of the Theory of Solids} (Cambridge University Press, 
Cambridge, 1972) 2nd ed.
\bibitem{RA} L. M. Roth and P. N. Argyres, in \textit{Semiconductors and Semimetals}, 
ed. R. K. Willardson and A. C. Beer (Academic Press, 1966) Vol. 1, Chap. 6, p159.\label{RA}
\bibitem{Ando} Y. Ando, J. Phys. Soc. Jpn. \textbf{82,} 102001 (2013).
\bibitem{Xiang} Z. J. Xiang, G. J. Ye, C. Shang, B. Lei, N. Z. Wang, K. S. Yang, D. Y. Liu, F. B. Meng, X. G. Luo, 
L. J. Zou, Z. Sun, Y. B. Zhang, and X. H. Chen, arXiv:1504.00125.
\bibitem{OpenMX} For the OpenMX package, code, pseudo-atomic basis functions and pseudopotentials, see http://www.openmx-square.org/.
\bibitem{Qiao} J. Qiao, X. Kong, Z.-X. Hu, F. Yang, and Q. Ji, Nat. Commun. {\bf 5,} 4475 (2014).
\bibitem{Cartz} L. Cartz, S. R. Strinivasa, R. J. Riedner, J. D. Jorgensen, and T. G. Worlton, J. Chem. Phys. {\bf 71,} 1718 (1979).
\end{thebibliography}
\end{document}